# Opportunistic Multi-Modal User Authentication for Health-Tracking IoT Wearables

Alexa Muratyan[1], William Cheung[1], Sayanton V. Dibbo[2], and Sudip Vhaduri[1]

[1] Fordham University, Bronx NY 10458, USA
[2] Dartmouth College, Hanover, NH 03755, USA
[1]`{amuratyan,wcheung5,svhaduri}@fordham.edu`, [2]`f0048vh@dartmouth.edu`

**Abstract.** With the advancement of technologies, market wearables are becoming increasingly popular with a range of services, including providing access to bank accounts, accessing cars, monitoring patients remotely, among several others. However, often these wearables collect various sensitive personal information of a user with no to limited authentication, e.g., knowledge-based external authentication techniques, such as PINs. While most of these external authentication techniques suffer from multiple limitations, including recall burden, human errors, or biases, researchers have started using various physiological and behavioral data, such as gait and heart rate, collected by the wearables to authenticate a wearable user implicitly with a limited accuracy due to sensing and computing constraints of wearables. In this work, we explore the usefulness of blood oxygen saturation $SpO_2$ values collected from the Oximeter device to distinguish a user from others. From a cohort of 25 subjects, we find that 92% of the cases $SpO_2$ can distinguish pairs of users. From detailed modeling and performance analysis, we observe that while $SpO_2$ alone can obtain an average accuracy of 0.69 and $F_1$ score of 0.69, the addition of *heart rate* (HR) can improve the average identification accuracy by 15% and $F_1$ score by 13%. These results show promise in using $SpO_2$ along with other biometrics to develop implicit continuous authentications for wearables.

## 1 Introduction

### 1.1 Motivation

With the explosion of the internet of things (IoT) and increased popularity of mobile networks [9], individuals' personal information is further exposed through the internet and IoT-connected wearables and other gadgets. According to a survey taken by over one thousand internet users conducted in September of 2020, one in five users have experienced an online account being compromised, and 70% of users have mentioned the hurdle to remember over ten passwords [7]. Thereby, knowledge-based authentications, such as passwords, PINs, and pattern locks, are not satisfactory for non-stop seamless IoT authentication. Therefore, biometric-based user authentication techniques are preferable to handle these issues.

However, most traditional biometrics, such as fingerprints [28], facial images [49], voice [22], breathing patterns [10], keystroke dynamics [24], and gait [16], are difficult, and nearly impossible, to adopt for tiny wearables with limited sensing and computing capabilities. While the IoT wearables are helping us with a wide range of services [6], these wearables also have the potential to authenticate a user implicitly and thereby secure the user's access to other IoT objects and accounts seamlessly and continuously [52].

While researchers have been relying on various biometrics collected by wearables, such as gait and breathing patterns, they have their limitations [27, 19]. For example, models developed for gait-based authentication do not work when a user is sedentary [40]. Therefore, there is a need for an authentication approach that can work continuously without a need for user input.

While various types of behavioral and physiological biometrics are already available in many market wearables, new types of data, such as oxygen saturation, collected continuously using the oxygen saturation ($SpO_2$) sensors and represent the percentage of oxygen-saturated hemoglobin compared to the total amount of hemoglobin in the blood, is becoming available to market wearables [4]. This personalized data could be valuable to identify an individual and thereby could be useful for implicit user authentication.

In addition to securing various IoT objects [38-40, 44, 15, 45], an implicit wearable-user authentication can be applicable to many other IoT supported services/sectors, e.g., health monitoring [32, 21, 47], well-being [34-36], stress monitoring [30], sleep quality improvement [41, 11], disease monitoring [31, 48, 29], and place discovery [33, 46, 37, 42, 43]. Thereby, it is essential to develop an implicit wearable user authentication system that can effortlessly validate a user's identity and secure the user's cyber-physical space in a non-stop fashion based on the data collected from the wearables.

### 1.2   Contributions

The main contribution of this work is to include oxygen saturation data in the development of an implicit wearable user authentication mechanism by using the oxygen saturation on its own or in combination with other standard biometrics, such as heart rate, used by other researchers [44]. We first test the usefulness of oxygen saturation data by attempting to distinguish an individual from others using the *Two-Sample T-tests* (Section 3.3). We find that 92% of the subject-pairs are significantly distinguishable based on their average oxygen saturation. This shows a promise to use oxygen saturation data to develop user authentication models. Next, we develop an authentication model using only oxygen saturation data, followed by a model using oxygen saturation with heart rate data (Section 3.6). While using oxygen saturation and heart rate together, we observe an improvement in average authentication accuracy of 0.80 compared to their solo performance (oxygen saturation: 0.69 and heart rate: 0.70) (Section 4.5). This shows the potential to develop and deploy a new implicit wearable user authentication using oxygen saturation and heart rate data, which could be further enhanced with additional biometrics.

## 2  Related Work

Compared to behavioral biometrics, such as gait, physiological biometrics, such as heart rate, is considered one of the most readily available biometric irrespective of the physical states, e.g., sedentary, and non-sedentary states, of a person. Fortunately, most of the market wearables are already equipped with photoplethysmogram (PPG) sensors to collect heart rate biometrics. Thereby, researchers began to develop various PPG-based authentication models.

Since solo PPG-based heart rate data-driven models can achieve an authentication success rate of around 0.90 [53], segregating data by activity helps to achieve an accuracy value around 0.95 [50]. But, the activity-based segregated models makes it difficult to find models for all possible activities due to the wide range of activities and their skewed distribution throughout the day. In addition to PPG sensors, researchers have been also utilizing electroencephalography [51], electrooculography [25], electrocardiogram [26], and electromiography [20] sensors to collect various types of biometrics from dedicated or prototypical wearables, which are not available in commonly found market-wearables. Thereby, the ultimate benefits of wearables cannot be fully utilized to develop an implicit and continuous IoT authentication.

However, the recent inclusion of FDA-approved electrocardiogram (ECG) sensors to Apple, Fitbit, and Samsung smartwatches has brought new opportunities for wearable devices and their implicit user authentication systems [3,5]. While the ECG sensors are innovative in providing more fine-grained user-specific information to boost the identification models further, these ECG sensors require a user to interact with the device, i.e., a user needs to complete an electric circuit to collect electrocardiogram data. Though experiments demonstrate newer and more powerful ECG sensors enable heart rate models to achieve around 0.99 accuracy [17], it will require more work to adapt for continuous and seamless authentication systems.

While it is relatively easy to develop and deploy single biometric-based user identification models, they usually suffer from low-performance [23]. In addition to the low performance of single biometric models, another flaw is that is once the biometric is compromised, there is nothing the user can do to unlock the device. Thereby, the user will not be able to access any information from that device [12]. As a result, multi-biometric models are emerging to be the most robust authentication strategy. Often, combining multiple complementary biometrics can achieve optimal performance. While researchers have been using multi-model wearable authentication using a heart rate, gait, and breathing hierarchy and has a $F_1$ score of 0.93, [13, 14], considering $SpO_2$ as an easily obtainable user-specific data could improve the performance of user authentication models.

Currently, $SpO_2$ is used in a multi-factor fingerprint authentication system. After the system performs a valid fingerprint check, it checks whether the heart rate and $SpO_2$ levels are at human levels preventing some spoof attacks [18]. But, that work does not utilize the capability of $SpO_2$ to uniquely identify an individual, which could improve multi-biometric user authentication systems' performance.

## 3   Approach

In this paper, we intend to demonstrate the importance and effectiveness of heart rate (*HR*) and oxygen saturation (*SpO$_2$*) data to identify wearable device users with the help of different machine learning models. Before we present the detailed analysis, we first introduce the datasets, pre-processing steps, usefulness of *SpO$_2$* data, feature engineering, and methods used in this work.

### 3.1   Wellue Dataset

We use the Wellue SleepU wrist-worn oxygen monitor to collect oxygen saturation (*SpO$_2$*) values and heart rate (*HR*) data. The data is gathered at a rate of one sample every four seconds. We collect data from 25 healthy subjects with average age $37 \pm 20.3$ years. Each subject wears the device continuously for 8 hours during his/her normal daily activity. The *SpO$_2$* and *HR* data are collected through the device's finger pulse oximeter sensor [2], stored locally in the wearable, and later transferred to a laptop using a USB connector.

### 3.2   Data Pre-Processing

Due to the device's extended wear time, there are missing entries where the sensor failed to record information. We define those missing entries as invalid data. To ensure our computations are accurate, we first clean the raw data to remove any invalid blocks.

Once the data is clean and all invalid data is removed, we segment the oxygen saturation data and its corresponding heart rate data into five different zones based on each subjects' demographics and maximum heart rate. Each zone can be categorized into a specific level of physical activity, ranging from very light to maximum activity, defined in Table 1.

**Table 1.** *Heart rate* zones [1]

| *HR* Zones | Ranges |
| --- | --- |
| 1 (very light) | 50% - 60% of max *HR* of an individual |
| 2 (light) | 60% - 70% of max *HR* of an individual |
| 3 (moderate) | 70% - 80% of max *HR* of an individual |
| 4 (intense) | 80% - 90% of max *HR* of an individual |
| 5 (very intense) | 90% - 100% of max *HR* of an individual |

In Section 3.3, we perform statistical tests to investigate whether the oxygen saturation data at different heart rate zones can be used to distinguish individuals.

Finally, we segment the continuous stream of heart rate and oxygen saturation information using fixed 40-second non-overlapping windows (i.e., 10-sample windows with one sample recorded every 4 seconds) to compute different types of statistical features to develop user authentication models (Section 3.4). In a window of continuous samples, we calculate a representative heart rate zone based on the majority voted zones of all samples in the window. We use this reference point, i.e., representative heart rate zone of a window, as an additional feature to our sets of statistical features.

### 3.3 Usefulness of Blood Oxygen Saturation

To determine the usefulness of the $SpO_2$ data, we use the *Two-Sample T-tests* in two-ways: comparing one subject with the other 24 subjects and comparing pairs of subjects. In both cases, we compare the average blood oxygen saturation values of two groups at a specific heart rate zone with a null hypothesis: "average oxygen saturation values of two groups are the same at a specific heart rate zone."

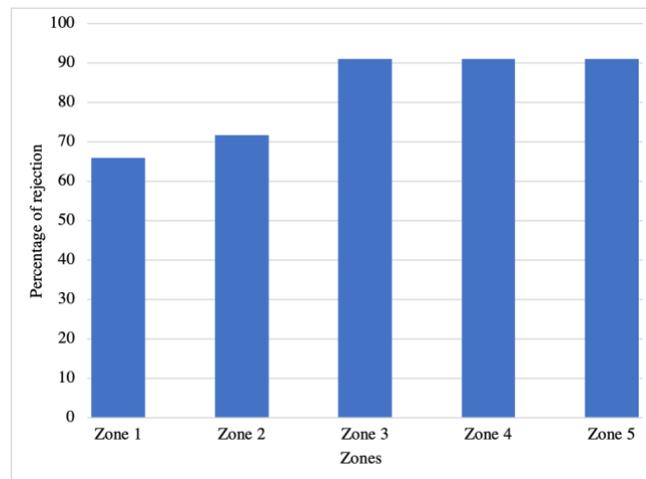

Fig. 1: *T-test* summary while distinguishing one subject from the rest of the 24 subjects based on $SpO_2$ data obtained at a specific heart rate zone

In the case of a rejection, we conclude that the average oxygen saturation values of the two groups are different at that specific heart rate zone, i.e., the reference point. Next, we aggregate the results obtained from all possible comparisons and normalize the rejection counts by the total number of comparisons to get an idea about the goodness (the higher, the better) of oxygen saturation to use as a person identification metric.

In Figure 1, we present our aggregated findings while comparing one subject to the other 24 subjects using the *Two-Sample T-tests* across the five heart rate zones. In the figure, we observe that from moderate to highly intensive heart rate zones, around 90% of the cases we reject the null hypothesis. However, lightly active heart rate zones have lower rejection percentages. Overall, we obtain an average rejection rate of 82% across all five heart rate zones while comparing one subject with the other 24 subjects at a specific zone. However, while comparing pairs of subjects and their average oxygen saturation values at a specific heart rate zone, we obtain an average rejection rate of 92%. These high rejection rates show a promise to use the oxygen saturation data to distinguish individuals.

### 3.4 Feature Computation

We compute the following sets of candidate features.

- Heart rate features: From each 10 sample (i.e., 40-second) window we compute 21 statistical features. They are mean, median, standard deviation, variance, coefficient of variance, range, coefficient of range, first quartile or $25^{th}$ percentile, third quartile or $75^{th}$ percentile, maximum, interquartile range, coefficient of interquartile, mean absolute deviation, median absolute deviation, energy, power, root mean square, root sum of squares, signal to noise ratio, skewness, and kurtosis.
- Oxygen saturation features: From each 10 sample (i.e., 40-second) window we compute the same 21 statistical features as listed in heart rate features.

Thereby, we obtain 21 statistical heart rate features, 21 statistical oxygen saturation features, and the window-level representative heart rate zone, i.e., the reference points, as an additional feature. Therefore, when we develop models from either heart rate or oxygen saturation data, we have 22 features in total. However, while using heart rate and oxygen saturation together, we have a total of 43 features.

### 3.5 Feature Selection

To identify the most influential features while training the binary classifiers, we use a two-level approach. In the first level, we remove highly correlated features. When a pair of features have a correlation value of higher than 0.9, we drop one of them; this way, we end up with a set of uncorrelated features. We use these uncorrelated features in our second-level.

In the second level, we try two different approaches using the sci-kit learn package: Principal Component Analysis (PCA) and Select the K-Best (SelectKBest).

PCA is a linear transformation algorithm, which is also a dimensionality-reduction method. Dimensionality is reduced by transforming a large set of variables into smaller ones. The entire feature set becomes condensed into vectors that best represent the data. SelectKBest is a form of univariate feature selection which works by selecting the best features based on univariate statistical tests. SelectKBest removes all but the K highest scoring features.

After analyzing both techniques, we find that PCA-based feature selection outperforms the SelectKBest-based features while keeping the feature count fixed. Once PCA was chosen, we test for the optimal feature count. We discovered 31 is an optimal feature count while using heart rate and oxygen data together, compared to 21 for the models using either heart rate or oxygen saturation data (details can be found in Section 4.4).

In the case of unary classifiers, the feature selection process is based on a different two-level approach. Similar to the binary classifiers, the first layer is passed through a correlation check with the same 0.9 correlation value used as a filtering threshold. Then, the second level focuses on selecting the features with lowest variance among the training set as influential features.

Similar to binary models, we find 31 as an optimal feature count while using heart rate and oxygen data together and 21 as optimal counts while using two types of data separately (details can be found in Section 4.4).

### 3.6 Methods

Based on the combination of the two types of data, i.e., oxygen saturation ($SpO_2$) and heart rate ($HR$) that we use to develop our authentication approaches, we define the following models:

- Heart rate data-driven model ($HR$ model)
- Oxygen saturation data-driven model ($SpO_2$ model)
- Heart rate and oxygen saturation data-driven model ($HRSpO_2$ model)

While developing the above models, we consider various classifiers, including random forest (RF), the $k$-nearest neighbor ($k$-NN), naive bayes (NB), and support vector machine (SVM) with binary and unary schemes. Compared to binary, unary models are available only for the SVM classifiers with radial basis function (RBF) and polynomial (Poly.) kernels.

## 4 User Authentication

Before presenting detailed evaluation of our models, we first present a list of performance measures, followed by training-testing set split, hyper-parameter optimization, and selection of an optimal number of features.

### 4.1 Performance Measures

To evaluate the performance of different modeling approaches, we consider the following measures:

*Accuracy (ACC)*, which is the fraction of predictions that are correct, i.e.,

$$ACC = \frac{TP+TN}{TP+FN+FP+TN} \qquad (1)$$

*Root Mean Square Error (RMSE)*, which is the square root of the sum of squares of the deviation from the prediction to the actual value. It is equivalent to the square root of the rate of misclassification, i.e.,

$$RMSE = \sqrt{\frac{FP+FN}{TP+FN+FP+TN}} \qquad (2)$$

***Genuine Rejection Rate (GRR)***, which is the fraction of invalid users rejected by an authentication system, or one minus the False Acceptance Rate (FAR), i.e.:

$$GRR = \frac{TN}{FP+TN} = 1 - FAR \qquad (3)$$

***Genuine Acceptance Rate (GAR)***, which is the fraction of valid users accepted by an authentication system or one minus the False Rejection Rate (FRR), i.e.:

$$GAR = \frac{TP}{TP+FN} = 1 - FRR \qquad (4)$$

***$F_1$ Score***, which is the measure of performance of an authentication system based on both its precision (positive predictive value) and recall (true positive rate) measures, i.e.:

$$F_1\ Score = 2\ (\frac{TP}{TP+FN} + \frac{TP}{TP+FP})^{-1} \qquad (5)$$

***Area Under the Curve - Receiver Operating Characteristic (AUC-ROC)***, which is the graphical relationship between FAR and FRR with the change of thresholds.
Where terminologies used in Equations 1, 2, 3, 4 and 5 have their usual meaning in machine learning, when classifying a subject using a feature set. Therefore, a desirable authentication system should have lower negative measures (i.e., RMSE, FAR, and FRR), but higher positive measures (i.e., ACC, $F_1$ Score, GRR, GAR, and AUC-ROC) of performance.

***Area***, which is the collective measure of performance based on accuracy, genuine rejection rate, genuine acceptance rate, $F_1$ score, and AUC-ROC measures. With each value ranging from zero to one, the area is defined as the pentagon shape of the connected performances for the binary cases and square shape (excluding AUC-ROC) for the unary case. The area computation of any classifier is normalized by dividing by the equivalent area in which there are perfect scores in all attributes, see Figures 4 and 5. This allows for classifiers to be comparable even between models, even among binary and unary models.

### 4.2   Training-Testing Set

In our binary model, we try to distinguish a valid user from the imposters. During the training-testing procedure, we follow the one-valid-user strategy. This means we train and test N unique models one-by-one, where in each iteration one of the N subjects is treated as the valid user. During each round of training-testing, use a 90%-10% train-test split with the imposter set having equal composition of the N-1 subject's data. The split is in sequential order which means the test data is the 10% that occurred at the end of the data collection. This best simulates a real use case scenario, as data is received by the model post training. Each subject is used once as a valid user's data with 25 iterations of training and testing in total.

### 4.3 Hyper-Parameter Optimization

We use the Sci-kit Learn library grid search package to find the most favorable hyper-parameter sets. The hyper-parameter optimization is performed separately for each iteration, and tested using various ranges of values. The different iterations of this approach resulted in similar values for the hyper-parameters, which are presented in Tables 2, 3, and 4.

### 4.4 Optimal Feature Count Determination

Before the feature selection process for the unary and binary models are executed, we first test the models through various feature count amounts. It is necessary to identify the most suitable feature count to avoid overfitting and underfitting. Overfitting is the case where the organization of the model is unreliable due to the fact that the model is learning too much information from the training data. Underfitting is the opposite, meaning the model learns too little information from the training data.

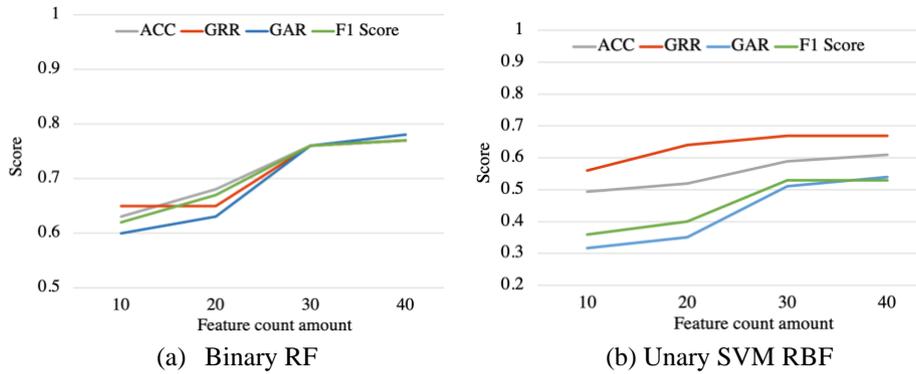

(a) Binary RF  (b) Unary SVM RBF

Fig. 2: Performance over feature count

To recognize the optimal count, we run the top performing binary model, RF, and top performing unary model, SVM RBF, through feature counts of 11, 21, 31 and 41. These specific numbers are due to the fact that we want to add the HR zones feature to the 10, 20, 30 and 40 features we found to be optimal. Figure 2a shows the preliminary analysis of performance of the binary HR$SpO_2$ across the mentioned feature counts. There is a considerable performance jump of eight percentage points in accuracy from 21 to 31 features, indicating increased learning. However, from 31 to 41 features there is not much gain in performance, which means it might start to over fit the data. Figure 2b shows the same analysis for unary models; we came to similar conclusions as we did for the binary case. For both the unary and binary models, 31 features is the optimal count. In the case of single biometric models, they do not have 31 features, so 21 are used.

### 4.5  Authentication Model Evaluation

In Tables 2, 3, and 4, we present the performance of the models using different biometric combinations and various classifiers. Heart rate data is commonly used in authentication models, so it is therefore used as the base metric in attempting to authenticate the user. Starting with the heart rate data-driven model (*HR* model), displayed in Table 2, we observe that the best classifier for the binary *HR* model is RF, as it provides an average ACC of 0.70 and AUC-ROC of 0.82. Compared to the binary, the *HR* model's unary classifiers present lower results, with SVM RBF having an average ACC of 0.58. This is foreseeable as the unary is not exposed to as much imposter training as the binary model is. To better the outcome of the *HR* model performance, which will implement the most accurate user authentication, we will add the oxygen saturation ($SpO_2$) data to the model. Combining the datasets will create a heart rate and oxygen saturation data-driven model (*$HRSpO_2$*).

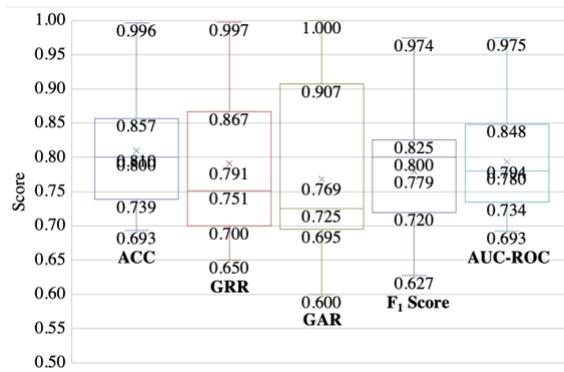

(a) *$HRSpO_2$* model with binary RF classifier

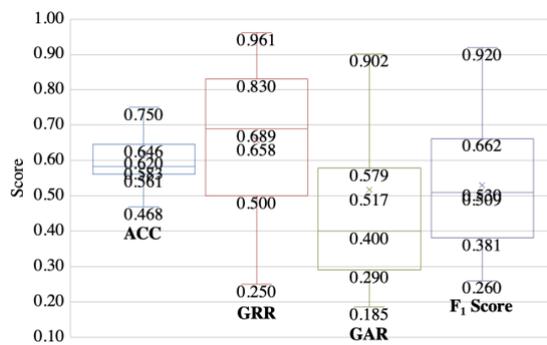

(b) *$HRSpO_2$* model with unary SVM RBF classifier

Fig. 3: Boxplots of different performance measures of the best binary and unary models

In Table 3, we observe that the top performing binary classifier, RF of the $SpO_2$ model, performs well, but in some cases is worse than the binary RF classifier in the *HR* model (Table 2). In comparison to the *HR* model, the $SpO_2$ model is 1% less accurate, GRR is reduced by 11%, and AUC-ROC is reduced by 23%. However, GAR and $F_1$ score are higher in the $SpO_2$, as GAR is 3% higher and $F_1$ score is 1.5% higher in reference to *HR*. As for the unary classifiers, the models both perform best with SVM RBF over SVM poly. Again, the *HR* model outperforms the $SpO_2$ model with 23% higher accuracy, 44% higher GRR, and 13% higher GAR. Although it may seem like the $SpO_2$ data is not useful, as it is less accurate than *HR* on its own, the $SpO_2$ compliments the *HR* data so that when combined, the heart rate and oxygen saturation data-driven model performs the best.

Table 2: *HR* models with average and standard deviation of performance measures

| **BINARY Model** | | | | | | | | |
|---|---|---|---|---|---|---|---|---|
| Classifier (parameters) | FC | ACC | RMSE | GRR | GAR | $F_1$ Score | AUC-ROC | Area |
| RF (n estimators = 50) | 21 | 0.70 (0.14) | 0.04 (0.01) | 0.79 (0.21) | 0.68 (0.16) | 0.68 (0.14) | 0.82 (0.10) | 0.52 |
| $k$-NN ($k$ = 5, minkowski distance) | 21 | 0.69 (0.14) | 0.04 (0.01) | 0.71 (0.21) | 0.67 (0.12) | 0.68 (0.13) | 0.69 (0.14) | 0.48 |
| NB | 21 | 0.64 (0.12) | 0.04 (0.01) | 0.79 (0.18) | 0.40 (0.23) | 0.48 (0.20) | 0.64 (0.12) | 0.38 |
| SVM (RBF kernel, $\gamma$ = 0.05, $C$ = 5) | 21 | 0.70 (0.14) | 0.04 (0.01) | 0.79 (0.17) | 0.61 (0.16) | 0.66 (0.16) | 0.69 (0.15) | 0.48 |
| SVM (Poly. kernel, $d$ = 3, $C$ = 12) | 21 | 0.67 (0.10) | 0.04 (0.01) | 0.80 (0.15) | 0.51 (0.19) | 0.59 (0.15) | 0.67 (0.10) | 0.43 |
| **UNARY Model** | | | | | | | | |
| SVM (RBF kernel, $\gamma$ = 0.05, $nu$ = 0.5) | 21 | 0.58 (0.10) | 0.05 (0.01) | 0.72 (0.24) | 0.42 (0.24) | 0.31 (0.18) | N/A | 0.28 |
| SVM (Poly. Kernel, $d$ = 1, $nu$ = 0.5) | 21 | 0.51 (0.07) | 0.05 (0.01) | 0.67 (0.28) | 0.32 (0.25) | 0.29 (0.17) | N/A | 0.22 |

Table 3: $SpO_2$ models with average and standard deviation of performance measures

| **BINARY Model** | | | | | | | | |
|---|---|---|---|---|---|---|---|---|
| Classifier (parameters) | FC | ACC | RMSE | GRR | GAR | $F_1$ Score | AUC-ROC | Area |
| RF (n estimators = 50) | 21 | 0.69 (0.11) | 0.04 (0.01) | 0.70 (0.25) | 0.70 (0.19) | 0.69 (0.12) | 0.63 (0.11) | 0.44 |
| $k$-NN ($k$ = 5, minkowski distance) | 21 | 0.64 (0.10) | 0.04 (0.01) | 0.61 (0.18) | 0.67 (0.16) | 0.64 (0.12) | 0.61 (0.10) | 0.37 |
| NB | 21 | 0.61 (0.08) | 0.04 (0.01) | 0.73 (0.19) | 0.50 (0.16) | 0.55 (0.12) | 0.61 (0.08) | 0.36 |
| SVM (RBF kernel, $\gamma$ = 0.05, $C$ = 5) | 21 | 0.66 (0.10) | 0.04 (0.01) | 0.67 (0.18) | 0.64 (0.10) | 0.65 (0.07) | 0.63 (0.08) | 0.40 |
| SVM (Poly. kernel, $d$ = 3, $C$ = 14) | 21 | 0.61 (0.08) | 0.04 (0.01) | 0.74 (0.17) | 0.49 (0.21) | 0.55 (0.15) | 0.61 (0.08) | 0.36 |
| **UNARY Model** | | | | | | | | |
| SVM (RBF kernel, $\gamma$ = 0.05, $nu$ = 0.5) | 21 | 0.47 (0.08) | 0.02 (0.00) | 0.50 (0.27) | 0.37 (0.25) | 0.42 (0.16) | N/A | 0.22 |

| SVM (Poly. Kernel, $d = 2$, $nu = 0.5$) | 21 | 0.47 (0.09) | 0.02 (0.00) | 0.49 (0.32) | 0.37 (0.27) | 0.36 (0.15) | N/A | 0.18 |

In Table 4, we observe that when adding the oxygen saturation data to the existing heart rate data, performance measures improve significantly. In the case of the best binary classifier (RF), we observe that the top $HRSpO_2$ model is 15% more accurate and has a 13% higher $F_1$ score compared to the best $SpO_2$ model. Similarly, the best $HRSpO_2$ model is 14% more accurate and has a 15% higher $F_1$ score, 3% higher GAR, and 17% higher area when compared with the *HR* model. These are substantial improvements made from the best *HR* model. Similar to the binary, the unary $HRSpO_2$ model shows promise over the unary *HR* model. The SVM RBF classifier has a 9% increase in ACC, a 21% increase in GAR, and a 71% rise in the $F_1$ score. These, too, are significant results in comparison to the unary *HR* classifiers.

Table 4: $HRSpO_2$ models with average and standard deviation of performance measures

| **BINARY Model** | | | | | | | | |
|---|---|---|---|---|---|---|---|---|
| Classifier (parameters) | FC | ACC | RMSE | GRR | GAR | $F_1$ Score | AUC-ROC | Area |
| RF (n estimators = 50) | 31 | 0.80 (0.09) | 0.04 (0.01) | 0.79 (0.14) | 0.77 (0.12) | 0.78 (0.09) | 0.78 (0.10) | 0.61 |
| $k$-NN ($k = 2$, minkowski distance) | 31 | 0.75 (0.01) | 0.04 (0.01) | 0.73 (0.17) | 0.68 (0.11) | 0.71 (0.09) | 0.69 (0.11) | 0.49 |
| NB | 31 | 0.66 (0.07) | 0.05 (0.01) | 0.84 (0.16) | 0.42 (0.16) | 0.49 (0.12) | 0.61 (0.08) | 0.38 |
| SVM (RBF kernel, $\gamma = 0.08$, $C = 3$) | 31 | 0.72 (0.13) | 0.04 (0.04) | 0.74 (0.17) | 0.71 (0.14) | 0.72 (0.13) | 0.71 (0.13) | 0.52 |
| SVM (Poly. kernel, $d = 4$, $C = 16$) | 31 | 0.69 (0.10) | 0.04 (0.01) | 0.77 (0.16) | 0.62 (0.13) | 0.67 (0.11) | 0.69 (0.11) | 0.50 |
| **UNARY Model** | | | | | | | | |
| SVM (RBF kernel, $\gamma = 0.05$, $nu = 0.5$) | 31 | 0.63 (0.08) | 0.05 (0.01) | 0.69 (0.27) | 0.51 (0.25) | 0.53 (0.16) | N/A | 0.37 |
| SVM (Poly. Kernel, $d = 1$, $nu = 0.75$) | 31 | 0.45 (0.10) | 0.06 (0.01) | 0.58 (0.32) | 0.28 (0.22) | 0.29 (0.17) | N/A | 0.19 |

In Figure 3a, we present five summarized values of different performance measures of the RF binary model. We observe that the median of each box plot is very similar to its average, which indicates there are not many outliers in the data. Additionally, we obtain tight interquartile ranges of about 0.15 for each of the performance measures, representing the consistency of the outcome. In the case of the unary SVM RBF model, shown in Figure 3b, interquartile ranges differ for each performance measure. ACC has a narrow interquartile range of 0.09, while GRR, GAR and $F_1$ score have a range of about 0.2.

The performance metrics are also presented in Figure 4 and Figure 5, through the form of spider graphs of the binary and unary models. Here we can visually understand the trade-offs of the various performances when choosing one algorithm over another in each model. In Tables 5, 6, and 7, we summarize GRR, GAR and Area to be used in the plot. GAR and GRR have meaning in security and represent the strength of the authentication system and therefore, are viewed as key measures. More specifically, GAR% tracks the accuracy of identifying if the valid user is actually valid. GRR% similarly tracks the accuracy of identifying if the invalid user is actually invalid. Although, we choose area as an overall metric to select the best model in certain cases GAR may hold higher significance.

Table 5: Relative performance-loss of *HR* models for a particular performance measure with respect to the best value of that measure. Negative signs are used to indicate a loss.

| **BINARY Model** | | | |
|---|---|---|---|
| Classifier (parameters) | GRR | GAR | Area |
| RF (n estimators = 50) | 0.79 (-1.25%) | 0.68 (0.00%) | 0.52 (0.00%) |
| *k*-NN (*k* = 5, minkowski distance) | 0.71 (-11.25%) | 0.67 (-1.47%) | 0.48 (-7.69%) |
| NB | 0.79 (-1.25%) | 0.40 (-41.18%) | 0.38 (-26.92%) |
| SVM (RBF kernel, $\gamma = 0.05$, $C = 5$) | 0.79 (-1.25%) | 0.61 (-10.29%) | 0.48 (-7.69%) |
| SVM (Poly. kernel, $d = 3$, $C = 12$) | 0.80 (0.00%) | 0.51 (-25.00%) | 0.43 (-17.31%) |
| **UNARY Model** | | | |
| SVM (RBF kernel, $\gamma = 0.05$, $nu = 0.5$) | 0.72 (0.00%) | 0.42 (0.00%) | 0.28 (0.00%) |
| SVM (Poly. Kernel, $d = 1$, $nu = 0.5$) | 0.67 (-6.94%) | 0.32 (-23.81%) | 0.22 (-21.43%) |

In Table 5 we can see that RF has the best overall performance with a minimal respective loss of 1.25% in GRR to NB. However, Random Forest makes up for it in terms of usability with 0.28 points higher in GAR. Oxygen saturation classifiers show a similar story in Table 6. RF is the top classifier, with a respective loss of 5.41% in GRR and no respective loss for the other metrics, indicating it is the top performer of the set. The most interesting point to notice is that once again, in *HRSpO₂*, RF outperforms other classifiers, as seen in Table 7. This solidifies the strength of the RF classifier as well as demonstrates the consistency between the models.

Table 6: Relative performance-loss of *SpO₂* models for a particular performance measure with respect to the best value of that measure. Negative signs are used to indicate a loss.

| **BINARY Model** | | | |
|---|---|---|---|
| Classifier (parameters) | GRR | GAR | Area |
| RF (n estimators = 50) | 0.70 (-5.41%) | 0.70 (0.00%) | 0.44 (0.00%) |
| *k*-NN (*k* = 5, minkowski distance) | 0.61 (-17.57%) | 0.67 (-4.29%) | 0.37 (-15.91%) |
| NB | 0.73 (-1.45%) | 0.50 (-28.57%) | 0.36 (-18.18%) |
| SVM (RBF kernel, $\gamma = 0.05$, $C = 5$) | 0.67 (-9.46%) | 0.64 (-8.57%) | 0.40 (-9.09%) |
| SVM (Poly. kernel, $d = 3$, $C = 14$) | 0.74 (0.00%) | 0.49 (-30.00%) | 0.36 (-18.18%) |
| **UNARY Model** | | | |
| SVM (RBF kernel, $\gamma = 0.05$, $nu = 0.5$) | 0.50 (0.00%) | 0.37 (0.00%) | 0.28 (0.00%) |
| SVM (Poly. Kernel, $d = 4$, $nu = 0.5$) | 0.49 (-2.00%) | 0.37 (0.00%) | 0.22 (-18.18%) |

We see the graphical representation of all the trade-offs in Figures 4 and 5. Even at a quick glance, it is evident that the classifiers of the binary *HRSpO₂* model, Figure 4, cover the majority of the area of the plot, which indicate that it is the optimal model. Choosing this algorithm is the obvious move, being that RF is the top performing classifier and has no respective loss for GAR or area. Now shifting to the unary plots, Figure 5, we observe that the SVM RBF classifier of the *HRSpO$_2$* model covers the greatest area. It is also important to note that of the Tables and Spider plot figures, the binary classifiers perform almost twice as well as the unary classifiers.

Table 7: Relative performance-loss of *HRSpO$_2$* models for a particular performance measure with respect to the best value of that measure. Negative signs are used to indicate a loss.

| BINARY Model | | | |
|---|---|---|---|
| Classifier (parameters) | GRR | GAR | Area |
| RF (n estimators = 50) | 0.79 (-5.95%) | 0.77 (0.00%) | 0.61 (0.00%) |
| *k*-NN (*k* = 2, minkowski distance) | 0.73 (-13.10%) | 0.68 (-11.69%) | 0.49 (-19.67%) |
| NB | 0.84 (0.00%) | 0.42 (-45.45%) | 0.38 (-37.70%) |
| SVM (RBF kernel, $\gamma$ = 0.08, *C* = 3) | 0.74 (-11.90%) | 0.71 (-7.79%) | 0.52 (-14.75%) |
| SVM (Poly. kernel, *d* = 4, *C* = 16) | 0.77 (-8.33%) | 0.62 (-19.48%) | 0.50 (-18.03%) |
| UNARY Model | | | |
| SVM (RBF kernel, $\gamma$ = 0.05, *nu* = 0.5) | 0.69 (0.00%) | 0.51 (0.00%) | 0.37 (0.00%) |
| SVM (Poly. Kernel, *d* = 1, *nu* = 0.75) | 0.58 (-15.94%) | 0.28 (-45.10%) | 0.19 (-48.65%) |

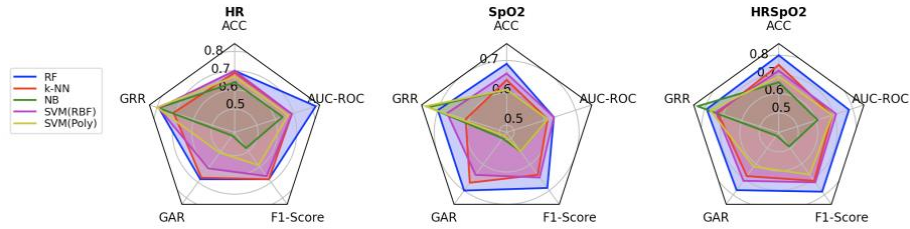

Fig. 4: Spider plot of the five binary model measures of performance

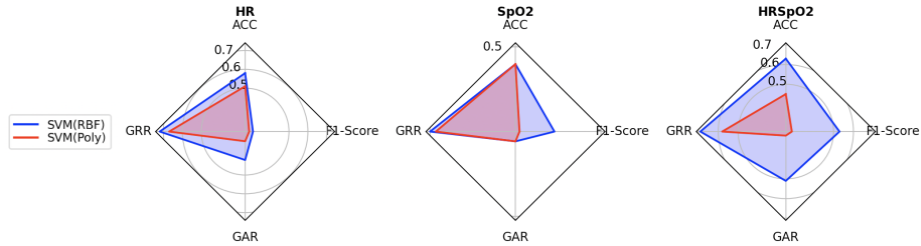

Fig. 5: Spider plot of the three unary model measures of performance

## 5   Conclusion and Future Work

To the best of our knowledge, this is the first attempt to use blood oxygen saturation value to identify an individual while developing an implicit and continuous wearable-device user authentication system. From our detailed analysis, we observe that oxygen saturation alone can provide around 0.69 average authentication accuracy using a random forest (RF) classification model. However, when combining with heart rate we can obtain around 0.80 authentication accuracy, i.e., 15% improvement. Interestingly, we also find that the heart rate alone provides a lower average authentication accuracy of 0.70, compared to the combined metrics. This shows the promise to develop a multi-model approach to authenticate a wearable device user implicitly and continuously using traditional heart rate biometric along with opportunistic oxygen saturation data.

As IoT technology continues to grow, more sensors are becoming available in common market wearables, such as accelerometers and photoplethysmogram (PPG) sensors [8]. Thereby, the $SpO_2$ is an opportunity to be used with other standard biometrics to increase the robustness and accuracy of multi-model biometric authentication systems. As mentioned before, some biometrics such as gait and breathing sounds are not always available. Compared to them, oxygen saturation is constantly available and provides a foundation piece for a multi-biometric scheme. However, this requires further long-term careful study with diverse sets of subjects over an extended period. We will also develop a multi-device authentication scheme, which will bring an opportunity to make more robust authentication scheme due to the increased popularity of IoT-based interconnected environments [44]. For example, an authentication mechanism can be developed by doing a fusion of Wellue data with similar or different types of data obtained from other wearables, such as Fitbit. In the future, we will investigate this multi-device implicit wearable-user authentication scheme.